\documentclass[prb,twocolumn,showpacs,superscriptaddress,floatfix]{revtex4}
\usepackage{graphicx,amsfonts,amssymb,amsmath,hyperref}

\newif\ifhyper
\hypertrue
\ifhyper
\hypersetup{
   citecolor = {green},
   colorlinks = {true}, 
   urlcolor = {blue} 
}
\fi

\newcommand{\rmi}{\mathrm{i}}
\newcommand{\ket}[1]{| #1 \rangle} 
\newcommand{\bra}[1]{\langle#1|} 
\newcommand{\braket}[2]{\langle#1|#2\rangle}

\begin{document}

\title{Exact spectrum of the Lipkin-Meshkov-Glick model in the thermodynamic limit and finite-size corrections}

\author{Pedro Ribeiro}
\email{ribeiro@lptmc.jussieu.fr}
\author{Julien Vidal}
\email{vidal@lptmc.jussieu.fr}
\author{R\'emy Mosseri}
\email{remy.mosseri@upmc.fr}
\affiliation{Laboratoire de Physique Th\'eorique de la Mati\`ere Condens\'ee, CNRS UMR 7600,
Universit\'e Pierre et Marie Curie, 4 Place Jussieu, 75252 Paris Cedex 05, France}

\begin{abstract}
The spectrum of the Lipkin-Meshkov-Glick model is exactly derived in the thermodynamic limit  by means of a spin-coherent-state formalism. In the first step, a classical  analysis allows one to distinguish between four distinct regions in the parameter space according to the nature of the singularities arising in the classical energy surface; these correspond to spectral critical points. The eigenfunctions are then analyzed more precisely in terms of the associated roots of the Majorana polynomial, leading to exact expressions for the density of states in the thermodynamic limit. Finite-size effects are also analyzed, leading in particular to logarithmic corrections near the singularities occurring in the spectrum. Finally, we also compute expectation values of the spin operators in a semiclassical analysis in order to illustrate some subtle effects occurring in one region of the parameter space.
\end{abstract}


\maketitle

%
%
\section{Introduction}
\label{sec:intro}
%
%

The Lipkin-Meshkov-Glick (LMG) model was proposed in 1965 to describe shape phase transitions in nuclei \cite{Lipkin65,Meshkov65,Glick65}. This model is often used to describe the magnetic properties of molecules such as ${\rm Mn}_{12}$ acetate \cite{Garanin98}. However, it  also captures the physics of interacting bosons in a double-well-like structure \cite{Turbiner88,Ulyanov92} and is thus relevant to (two-mode) Bose-Einstein condensates \cite{Cirac98} as well as Josephson junctions. It has also been recently used in optical cavity quantum electrodynamics in its dissipative version \cite{Morrison08_1,Morrison08_2} for studying the decoherence of a single spin coupled to a spin bath \cite{Hamdouni07,Quan07} or  quench dynamics \cite{Das06}. Note also that, in recent years, the entanglement properties of its ground state \cite{Vidal04_1,Vidal04_2,Vidal04_3,Latorre05_2, Unanyan05_1,Barthel06_2,Vidal07,Kwok08,Orus08_2,Cui08} as well as the finite-size behavior \cite{Dusuel04_3,Dusuel05_2,Leyvraz05,Rosensteel08} have focused much attention on this model.

 An exact solution of this model has been derived \cite{Pan99,Links03_1,Ortiz05}, but it requires the solution of Bethe-like equations, which is more costly in terms of computational effort than exact diagonalization. Although the low-energy physics of the model has been widely studied through different approaches (variational \cite{Lipkin65,Botet82,Botet83}, bosonization \cite{Dusuel04_3,Dzhioev04,Chen06}, and coherent states \cite{Kuriyama03,Chen06}), its high-energy properties have only been very recently investigated numerically \cite{Heiss05,Heiss06,Castanos06} and several interesting features have been revealed. 
More precisely, for special values of the energy, the spectrum has been shown to display singularities which are reminiscent of the critical point responsible for the well-known quantum phase transition at zero temperature. 

In a recent Letter \cite{Ribeiro07}, we proposed a theoretical framework which allows for an exact computation, in the thermodynamic limit, of the LMG model spectrum for the whole range of parameters and leads to a precise description and understanding of these so-called exceptional points. The present paper is an extension of that work, in which we will detail some of its main results and extend the analysis along several directions (relation to the semiclassical treatment, first-order finite-size corrections and expectation values of observables).  

The paper is organized as follows. In Sec.~\ref{sec:model}, we introduce the LMG model and the spin-coherent-state formalism \cite{Klauder85}, which is the key ingredient of our approach. We then derive the classical energy surface \cite{Castanos06}, whose extrema lead to a qualitative phase diagram; these extrema are related to the density-of-states singularities. 
In a second step, most importantly, we analyze this phase diagram quantitatively.
In Sec.~\ref{sec:Majorana}, we introduce the Majorana polynomial and map the time-independent 
Schr\"{o}dinger equation onto a first-order nonlinear (Riccatti-like) differential equation. In Sec.~\ref{sec:thermo}, we give solutions of this equation in the thermodynamic limit. This leads to simple expressions of the density of states in the whole phase diagram. In Sec.~\ref{sec:finite}, 
we go beyond this limit and compute the leading finite-size corrections to the density of states. Finally, in Sec.~\ref{sec:observables}, we compute  the expectation values (throughout the spectrum) of some spin observables, paying  particular attention to one region for which spectral subtleties prevent us from using the Hellmann-Feynman theorem.
Some technical details are given in the Appendix.
%
%
\section{Coherent-state representation and classical energy surface}
\label{sec:model}
%
%

%
%
\subsection{Lipkin-Meshkov-Glick model}
%
%

The LMG model describes a set of $N$ spin-$\frac{1}{2}$ particles mutually interacting through an (anistropic) $XY$-like Hamiltonian and coupled to an external transverse magnetic field $h$. The Hamiltonian of this system can be expressed in terms of the total spin operators 
$S_{\alpha}=\sum_{i=1}^N \sigma_{\alpha}^{i}/2$ where the $\sigma_{\alpha}$'s are the Pauli matrices:
%
%
\begin{equation}
\label{eq:hamiltonian}
H=-\frac{1}{N} \big(\gamma_x S_x^2 + \gamma_y S_y^2  \big) - h \: S_z.
\end{equation}
%
%

In the following, for simplicity, we only consider the maximum spin sector $s=N/2$ with $N$ even.
Given the symmetry of the spectrum of $H$, we focus on the parameter range  $h \geqslant 0$;  
$ |\gamma_y| \leqslant \gamma_x$. Note also that $\big[ H, {\bf S}^2 \big]=0$ and 
$\big[ H, \mathrm{e}^{\mathrm{i} \pi  (S_z - s)} \big]=0$ (spin-flip symmetry).  In the standard eigenbasis 
$\{|s,m\rangle\}$ of ${\bf S}^2$ and $S_z$, this latter symmetry implies that odd- and even-$m$ states decouple.
In the thermodynamic limit, both subspaces are isospectral so that we further limit the following analysis to the $(s+1)$-dimensional sector with $m$ even. It is known that $H$ exhibits a quantum phase transition for $h=\gamma_x$ or $h=\gamma_y$.
%
%
\subsection{Coherent-state representation of the spin operators}
%
%

To determine the spectrum of the Hamiltonian $H$, it is convenient to use a spin-coherent-state representation \cite{Klauder85}. Let us denote by $\{|s,m\rangle\}$ the standard eigenbasis of $\big\{{\bf S}^2, S_z\big\}$ with eigenvalues $s(s+1)$ and $m$, respectively. The unnormalized spin coherent state $\ket{\alpha }$ is then defined as 
 %
%
\begin{equation}
\label{eq:coherent_states}
\ket{\alpha } =   e^{\bar{\alpha} S_+} \ket{s,-s} .
\end{equation}
%
%
The scalar product of two such states is 
%
%
\begin{equation}
\label{eq:inner_product_coherent_states}
\braket{\alpha'}{\alpha} =   (1+\bar{\alpha} \alpha')^{2s},
\end{equation}
%
%
where $\bar{\alpha}$ is the complex conjugate of $\alpha$. These coherent states obey the following closure relation:
%
%
\begin{equation}
\label{eq:closure_relation_coherent_states}
\int \frac{\mathrm{d}\bar{\alpha }  \mathrm{d}\alpha }{\pi }\: \frac{(2 s + 1)}{\left(1+\bar{\alpha }\alpha   \right)^{ 2} } \frac{\ket{\alpha } \bra{\alpha }}{\braket{\alpha }{\alpha }} = 1,
\end{equation}
%
%
where $\int \mathrm{d}\bar{\alpha }\mathrm{d}\alpha=\int \mathrm{d} \text{Re} (\alpha)  \ \mathrm{d} \text{Im} (\alpha)$. 
In this representation, a quantum state $\Psi(\alpha)=\braket{\alpha}{\Psi}$ is a polynomial in $\alpha$, and the action of the spin operators on $\Psi$ translates into differential operators:
%
\begin{eqnarray} 
S_+ &=& 2s \alpha-\alpha^2 \partial_\alpha, \label{eq:coherent_spin1} \\
S_-  &=& \partial_\alpha, \label{eq:coherent_spin2}\\ 
S_z &=& -s+ \alpha \partial_\alpha, \label{eq:coherent_spin3}
\end{eqnarray}
%
%
where $S_\pm=S_x \pm \rmi S_y$. We shall discuss below the representation of $\Psi(\alpha)$ in terms of its zeros (the Majorana representation).

%
\subsection{Classical energy surface}
%

In the thermodynamic limit, a variational description of the ground state \cite{Lipkin65,Botet82,Botet83}, built with respect to the $\ket{\alpha } $ states, leads to the dominant behavior  of the model and, in particular, the location of the quantum phase transition. The latter can obtained from an analysis of the minima of the variational energy $\mathrm{H}_{0}$:
%
%
\begin{eqnarray}
\label{eq:mean_field}
\mathrm{H}_{0} (\bar{\alpha }, \alpha )  &=&
 \lim_{s\to \infty } \, \frac{1}{s} \frac{\bra{\alpha}H | \alpha \rangle}{\braket{\alpha}{\alpha}} ,\\
 &=&  
 \frac{2\left(1-\alpha ^2 \bar{\alpha }^2\right)h-\left(\alpha +\bar{\alpha }\right)^2 \gamma _x+\left(\alpha -\bar{\alpha }\right)^2 \gamma_y}{2 \left(1+\alpha  \bar{\alpha }\right)^2}. \nonumber \\
\end{eqnarray}
%
%

Note that, in this limit, a classical spin description is valid, such that the correspondence between a state $\ket{\alpha}$ and a classical vector is simply obtained via a stereographic map from the complex plane onto the $\mathcal{S}^2$ sphere [with $\alpha = \mathrm{e}^{\mathrm{i} \theta} \tan(\phi/2)$], leading to the parametrization
%
%
\begin{equation}
{\bf S} ={N \over 2} (\sin \theta \cos \phi, \sin \theta \sin \phi, \cos \theta).
\end{equation}

Here we shall first be interested in the geometrical properties of the whole classical energy surface $\mathrm{H}_{0} (\bar{\alpha }, \alpha )$. Its extrema, obtained by imposing $\partial _{\bar{\alpha }} \mathrm{H}_{0} =  \partial _{\alpha }\mathrm{H}_{0} = 0$, are given in Table \ref{tab:mean_field_extremum} together with the corresponding energy. When one further imposes that $\alpha$ and $\bar{\alpha }$ be complex conjugate, the configuration space (spanned by the Hamiltonian parameters) is split into distinct regions characterized by the number of extrema and saddle points in $\mathrm{H}_{0} (\bar{\alpha }, \alpha )$.

This phase diagram coincides with that derived from the analysis of density of states singularities, as done in the next section. We shall describe below how far the classical analysis can help in understanding the spectral results. Note that a related analysis of the classical energy surface, including comparisons to numerically derived spectra, has already been proposed by Casta{\~ n}os {\it et~al.} \cite{Castanos06} in terms of the $(\theta,\phi)$ angles instead of the present ($\bar{\alpha }, \alpha$).
%
%
\begin{table}[h]
\center
\begin{tabular}{|c|c|c|}
\hline
$\alpha$  &
$\bar{\alpha }$ & 
$\mathrm{H}_{0} $ \\ \hline
$0 $ &
$0$ & 
$h$ \\  \hline
$-\Big(\frac{-h-\gamma _x}{h-\gamma _x}\Big)^{1/2}$ &
$ -\Big(\frac{-h-\gamma _x}{h-\gamma _x}\Big)^{1/2}$  & 
$-\frac{h^2+\gamma _x^2}{2 \gamma _x} $ \\ \hline
$\Big(\frac{-h-\gamma _x}{h-\gamma _x}\Big)^{1/2} $ &
$ \Big(\frac{-h-\gamma _x}{h-\gamma _x}\Big)^{1/2} $ &
$ -\frac{h^2+\gamma _x^2}{2 \gamma _x} $\\  \hline
$ -\Big(\frac{h+\gamma _y}{h-\gamma _y}\Big)^{1/2} $ &
$ \Big(\frac{h+\gamma _y}{h-\gamma _y}\Big)^{1/2} $ &
$ -\frac{h^2+\gamma _y^2}{2 \gamma _y} $\\ \hline
$\Big(\frac{h+\gamma _y}{h-\gamma _y}\Big)^{1/2} $ &
$ -\Big(\frac{h+\gamma _y}{h-\gamma _y}\Big)^{1/2}  $ &
$ -\frac{h^2+\gamma _y^2}{2 \gamma _y}$\\ \hline
$\infty$  &
$ \infty  $
& $-h$\\  \hline  
\end{tabular}
\caption{\label{tab:mean_field_extremum} Extrema of the energy surface $\mathrm{H}_{0}$.}
\end{table}
%
%

%
%
\subsection{Classical description of the phase diagram}
\label{subsec:PD}
%
%

The zero-temperature phase diagram of the LMG model is usually discussed in terms of its ground-state properties. In this case, only two phases are distinguished \cite{Lipkin65,Botet83,Dusuel05_2}. 
For $h>\gamma_x$ (symmetric phase), the ground state is unique and $\lim_{s\rightarrow \infty}\left\langle S_z\right\rangle/s=1$, whereas for $h <\gamma_x$ (broken phase), the ground state is twofold degenerate and  $\lim_{s\rightarrow \infty}\left\langle S_z\right\rangle/s=h/\gamma_x$. Note that the degeneracy in the broken phase arises only in the thermodynamic limit, where the gap between the ground and first excited states vanishes exponentially with $s$. 
The quantum phase transition at $h =\gamma_x$ is of second order and characterized by mean-field critical exponents \cite{Botet83} as well as nontrivial finite-size scaling behavior \cite{Dusuel04_3,Dusuel05_2,Leyvraz05}. 

\begin{figure}[t]
 \centering
\includegraphics[width=8cm]{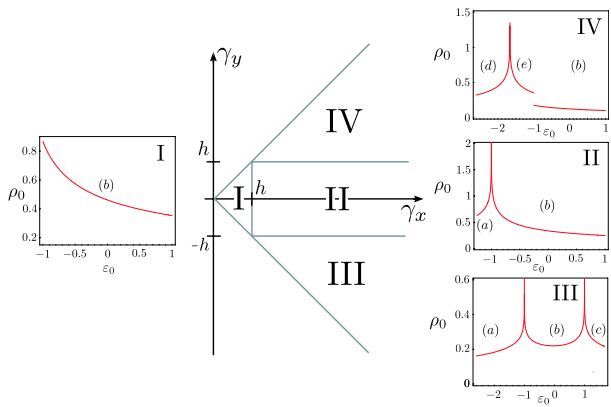}
\caption{\label{fig:phase_diagram} 
Phase diagram in the ($\gamma_x,\gamma_y$) plane at fixed $h>0$ and typical density of states for ($\gamma_x, \gamma_y, h$) equal to I: (1/2, 1/3, 1), II: (2, 1/2, 1), III: (5, -3, 1), and IV: (5, 3, 1).}
\end{figure}

We have shown \cite{Ribeiro07} that, when considering the full spectrum, four different zones arise instead of two, corresponding to a splitting of the broken phase region into three distinct parts characterized by different singularities in the density of states  (see Fig.~\ref{fig:phase_diagram}). 
%
%
%
%

Note that such singularities have already been pointed out in the numerical study of the special case $\gamma_x=-\gamma_y$ \cite{Heiss05,Heiss06} and were called ``exceptional points.''
We emphasize in the present study that \emph{these exceptional points are associated with saddle points of the energy surface.} Of course, the absolute minimum (maximum) gives the  lower (upper) bound of the spectrum. Note that these bounds may be degenerate.

In the thermodynamic limit, to a given energy in the spectrum corresponds a level set on $\mathrm{H}_{0} (\bar{\alpha }, \alpha )$. At that energy, the Husimi function local maxima (defined in the next section) are known to concentrate along this level set, which forms the classical orbit. Singularities of the surface (maxima, minima, or saddle points) translate into singularities of the level sets (a main ingredient in Morse surface theory). This, in turn, affects the density of states computation, as illustrated in the next section, and explains why the singularities in the $\mathrm{H}_{0} (\bar{\alpha }, \alpha )$ surface and in the density of states are in close correspondance.

\begin{figure}[t]
 \centering
\includegraphics[width= 8cm]{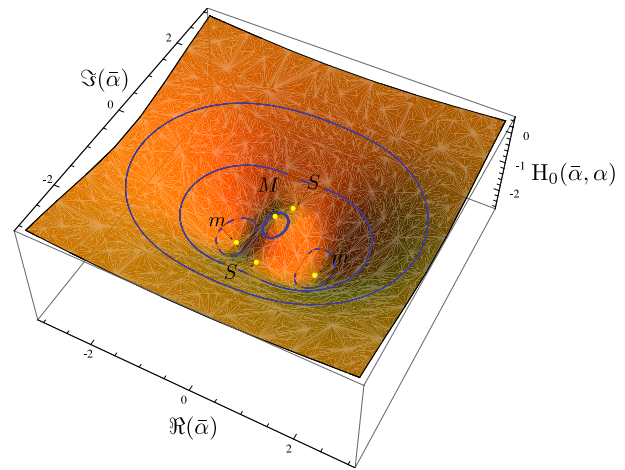}
\caption{\label{fig:mean_field_surface}  Typical classical energy surface in zone IV $( \gamma_x=5, \gamma_y=3, h=-1 )$, containing several critical points: two minimal points ($m$); two saddle points ($S$);  one local maximum ($M$). It also contains a global maximum, outside the range of this plot. The level curves of $\mathrm{H}_{0}$ (classical trajectories) are plotted in blue. }
\end{figure}

As an illustration, we display in Fig.~\ref{fig:mean_field_surface} the classical energy surface for 
$(\gamma_x=5,\gamma_y=3,h=-1)$, which is precisely the point of zone IV whose density of states is shown in 
Fig.~\ref{fig:phase_diagram}. As can be seen, the density of states contains two different types of singular points, being the locus of either a divergence or discontinuity. The analysis of the classicalenergy surface allows one to qualitatively understand all these features. Indeed, it contains two absolute minima (denoted $m$) which provide the lower bound of the spectrum (twofold-degenerate ground-state energy); two saddle points (denoted $s$) corresponding to the singular behavior of density of states; 
one local maximum (denoted $M$) which is associated with the discontinuity, and one absolute maximum, not shown here, giving the upper bound of the spectrum. 
 
%

The same geometrical analysis can be performed throughout the configuration space. A typical classical surface in zone I displays one minimum and one maximum, which, respectively, signal the lower and upper edges of the spectrum. A zone-II surface has two absolute minima (corresponding to the broken-phase degenerate ground states), a saddle point (corresponding to the density-of-states singularity), and one maximum (the upper spectrum edge). Finally, a generic zone-III surface has (again) two absolute minima, two saddle points (corresponding to the two singularities in the spectrum, arising at different energies), and two absolute maxima (corresponding to a degenerate upper state).
Note that, when displayed on the sphere, one recovers the standard result for surfaces singularities, which states that the number of maxima plus the number of minima minus the number of saddle points equals the genus of the sphere, {\it i.~e.}, 2.

Thus, the analysis of the classical energy surface allows us to qualitatively describe the phase diagram shown in Fig.~\ref{fig:phase_diagram}. However, it does not give any quantitative information concerning the density of states. The aim of what follows is to develop a reliable method to exactly compute the full spectrum of the LMG model.

%
%
\section{Majorana representation and spectrum}
\label{sec:Majorana}
%
%

%
%
\subsection{Majorana polynomial and Majorana sphere}
%
%

The first step consists in analysing the eigenstates in the spin-coherent-state formalism. Any $\ket{\Psi}$ can be represented by its Majorana polynomial \cite{Majorana32} defined as
%
%
\begin{eqnarray}
\label{eq:polynome} 
\hspace{-5mm} \Psi(\alpha)&=&\braket{\alpha}{\Psi} \\
&=& \sum _{m=-s}^s  \sqrt{\frac{(2 s)!}{(s-m)! (m+s)!}} \braket{s,m}{\Psi}  \alpha^{m+s} \\
&=&  C \prod_{k = 1}^{d} \left(  \alpha - \alpha_k  \right) ,
\end{eqnarray}
%
%
where $d \leqslant 2s$ is the degree of this polynomial in $\alpha$ ($d=2s$ for a generic state). The roots $\alpha_k$ of  $\Psi(\alpha)$ fully characterize the normalized quantum state $\ket{\Psi}$ up to a global phase. 

It may be more convenient to represent such a state $\ket{\Psi}$ on the so-called Majorana sphere, which can be seen as a generalization of the celebrated Bloch sphere used for spin-$\frac{1}{2}$ states. To do so, one first complements 
the $d$ roots  $\Psi(\alpha)$ with $2s-d$ roots at infinity in the complex plane. Next, the resulting set of $2s$ complex numbers $\alpha_k$ is mapped onto $2s$ points on the unit sphere by an inverse stereographic map. For instance, the basis states $|s,m\rangle$ are represented by 
$s-m$ points on the north pole and $s+m$ points on the south pole. Less trivial examples can be found in Fig.~\ref{fig:Majsphere} for eigenstates of $H$ in the zone III.

Let us also introduce  $G(\alpha)$, the logarithmic derivative of $\Psi(\alpha)$
%
%
\begin{equation}
\label{G_definition}
G(\alpha) = \frac{1}{2s} \partial_\alpha \log \Psi(\alpha) = \frac{1}{2s}  \sum_{k=1}^{2s} \frac{1}{\alpha-\alpha_k}.
\end{equation}
%
%
The $1/{2s}$ factor is here to ensure that $G$ is well behaved at the (infinite-$s$) thermodynamic limit.
Let us also define the Husimi function associated with a general state $\Psi(\alpha)$, 
%
%
\begin{eqnarray}
\label{eq:Husimi}
\hspace{-5mm} W_{\Psi}(\bar{\alpha},\alpha) &=& \frac{\braket{\alpha}{\Psi} \braket{\Psi}{\alpha}}{\braket{\alpha}{\alpha}} ,\\
&=& {\rm e}^{ 2s \left[ \int^{\alpha} G(\alpha') {\rm d}\alpha'  +  \int^{\bar{\alpha}} \bar{G}(\bar{\alpha}') {\rm d}\bar{\alpha}'  - \log (1+ \bar{\alpha} \alpha)  \right]}.
\end{eqnarray}
%
%
We shall further need to locate the Husimi function extrema,  which are easily found to satisfy 
%
%
\begin{equation}
\label{eq:Husimi_ext}
G(\alpha) = \frac{\bar{\alpha}}{1+\bar{\alpha} \alpha } .
\end{equation}
%
As explained above, for the Hamiltonian eigenstates, these maxima converge at the thermodynamic limit, toward the semiclassical orbits, which are the level sets of the classical energy surface $\mathrm{H}_{0}$.

\subsection{From Schr\"{o}dinger to Riccati}
%
%

Let us now write the time-independent Schr\"{o}dinger equation $H |\Psi \rangle =E |\Psi \rangle$ in the coherent-state representation. Using relations (\ref{eq:coherent_spin1}), (\ref{eq:coherent_spin2}), and (\ref{eq:coherent_spin3}), one transforms the Schr\"{o}dinger equation into the linear differential equation
%
%
\begin{equation}
\label{eq:Maj_Schrodinger}
\bigg[\frac{P_2(\alpha)}{(2s)^2} \partial^2_{\alpha}+ \frac{P_1(\alpha)}{2s} \partial_{\alpha} + P_0(\alpha ) \bigg] \Psi(\alpha) = \varepsilon \Psi(\alpha),
\end{equation}
%
%
where $\varepsilon=E/s$ and
%
%
\begin{eqnarray}
\label{eq:G_Schrodinger_polynomes} 
\hspace{-5mm} P_0(\alpha ) &=&  \frac{1}{4s} \Big[ \alpha^2  (2s-1)(\gamma _y - \gamma _x )    -  \gamma _x - \gamma_y \Big] +h, \\ 
\hspace{-5mm} P_1(\alpha) &=&  \alpha \bigg\{\frac{2s-1}{2s}  \Big[ \alpha^2(\gamma_x-\gamma_y)-\gamma_x-\gamma_y  \Big] -2h  \bigg\}, \\ 
\hspace{-5mm} P_2(\alpha) &=& - \frac{1}{2} \left[\left(\alpha^2-1\right)^2 \gamma _x-\left(\alpha^2+1\right)^2 \gamma _y \right]. 
\end{eqnarray}
%
%
The next step consists in converting the linear second-order differential equation (\ref{eq:Maj_Schrodinger}) for $\Psi$  into a nonlinear first-order differential equation for its logarithmic derivative $G(\alpha)$, which  satisfies the following Riccati-like equation
%
%
\begin{equation}
\label{eq:G_Schrodinger}
P_2(\alpha) \left[ \frac{G'(\alpha)}{2 s} + G^2(\alpha) \right] + P_1(\alpha ) G(\alpha) +  P_0(\alpha) = \varepsilon.
\end{equation}
%

\subsection{Density of states and poles of $G$}

\begin{figure}[t]
 \centering
\includegraphics[width=8.5cm]{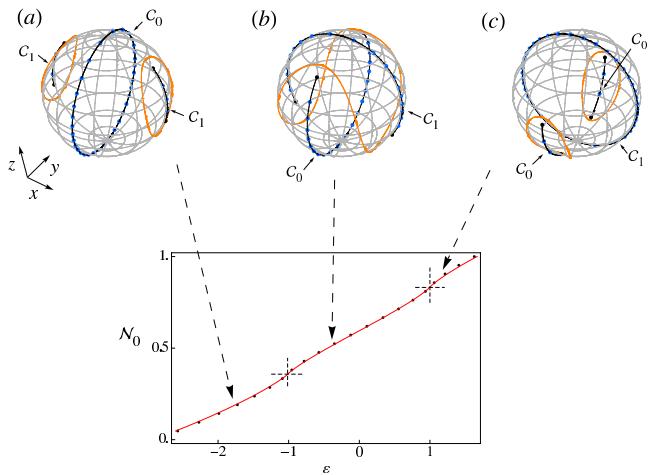}
\caption{ \label{fig:Majsphere}  Upper part: representation of the poles of $G$ on the Majorana sphere (blue dots) for three typical eigenstates computed for $h=1, \gamma_x=5, \gamma_y=-3$ and $s=20$ (zone III in Fig.~\ref{fig:phase_diagram}). Black lines correspond to the $G_0$ branch cuts $\mathcal{C}_0$ and 
$\mathcal{C}_1$; orange lines correspond to the classical orbits. 
Lower part: numerical (black dots $s=20$) versus analytical (red line $s=\infty$) integrated density of states. 
The two crosses indicate the singularities of the density of states ${N}^{\rm III}_0(-h)$ and ${N}^{\rm III}_0(h)$ [Eqs.~(\ref{eq:NII}) and (\ref{eq:NIII}), respectively] in  the thermodynamic limit.}
\end{figure}

The density of states is then obtained from the analysis of the poles of the function $G$. To illustrate the poles location,  several typical states are displayed in Fig.~\ref{fig:Majsphere} on the Majorana sphere. Each dot represents one pole of $G$, {\it i.~e.}, one Majorana zero $\alpha_k$, which is mapped from the complex plane to the sphere by an  inverse stereographic projection.

The cornerstone of this study is that, for the LMG model, the $\alpha_k$'s spread over two curves $\mathcal{C}_0$ and $\mathcal{C}_1$ in the complex plane. In addition, the $n$th excited state of $H$ has $2n$ poles on $\mathcal{C}_1$ and $2(s-n)$ on $\mathcal{C}_0$ (thus defining  both curves). This remarkable property stems mainly from existing maps (which may differ between parameter space regions) between the LMG model and the problem of a particle in an effective one-dimensional potential (see the Appendix). In the latter case, the oscillation theorem indexes the excited states  by the number of wave-function nodes on the real axis. This leads here to (at least one set of) zeros lying on simple lines in the complex plane, where the pole density varies monotoneously with energy.

%
%
%
%

Let us consider the normalized integrated density of states, $\mathcal{N}(\varepsilon) \in[0,1]$. We shall enumerate by $n$ the eigenstates of increasing energy, starting from $n=0$ for the ground state to $n=s$ for the highest-energy state. The special location of the $G$ poles leads to a simple relation between $\mathcal{N}(\varepsilon)$ and $p$, the number of poles lying in $\mathcal{C}_1$, which reads
%
%
\begin{eqnarray}
\label{eq:int_of_G}
\mathcal{N}(\varepsilon)&=&\frac{n+1}{s+1}=\frac{1}{s+1}\left(1+\frac{p}{2} \right) \\
&=& \frac{1}{s+1}\bigg[1+ \frac{s}{2 \rmi \pi} \oint_{\widetilde{\mathcal{C}}_1} G(\alpha)  \:  {\rm d} \alpha\bigg],
\end{eqnarray}
%
%
where $\widetilde{\mathcal{C}}_1$ is a contour that surrounds $\mathcal{C}_1$ and oriented such that $\mathcal{N}\geq 0$. For the sake of simplicity, we shall further consider the density of poles in $\mathcal{C}_1$, called $\mathcal{I} \in[0,1]$, which simply reads
%
%
\begin{equation}
\label{eq:int_of_G}
\mathcal{I}(\varepsilon)=\frac{p}{2s}=\frac{1}{2 \rmi \pi} \oint_{\widetilde{\mathcal{C}}_1} G(\alpha)  \:  {\rm d} \alpha.
\end{equation}
%
In general, Eqs.~(\ref{eq:int_of_G}) and (\ref{eq:G_Schrodinger}) cannot be exactly solved for arbitrary $s$. 
The main goal of this paper is to solve these in the thermodynamic limit ($s \rightarrow \infty$) and to capture the leading finite-size corrections in a $1/s$ expansion.

%
%
\section{Thermodynamic limit}
\label{sec:thermo}
%
%
%

%
%
\subsubsection {Leading-order expansion for $G$}
%

Let us assume that $G$ and $\varepsilon$ can be expanded in the form
%
%
\begin{equation}
\label{eq:one_over_s_development}
G = \sum_{i \in \mathbb{N}} \frac{G_i}{s^i}, \quad  \varepsilon = \sum_{i \in \mathbb{N}} \frac{\varepsilon_i}{s^i}.
\end{equation}
%
%
At leading order $(1/s)^0$, Eq.~(\ref{eq:G_Schrodinger}) becomes a second-order polynomial equation for  
$G_0$ whose solutions are
%
%
\begin{equation}
\label{eq:G_0_sol}
G^\pm_0 (\alpha) = \frac{\alpha \big[\alpha^2(\gamma_y-\gamma_x)+\gamma_x+\gamma_y +2h\big]  \pm \sqrt{2 Q(\alpha)}}{2 P_2(\alpha)},
\end{equation}
%
%
where 
%
%
\begin{eqnarray}
Q(\alpha)&=&\kappa  \left(\alpha^2-\mathit{r}_-^2\right) \left(\alpha ^2-\mathit{r}_+^2\right), \\
\kappa & = & -\left(\gamma _x-\gamma _y\right) \left(h+\varepsilon _0\right), \\
\label{eq:rpm}
\mathit{r}_{\pm}& = & (-\kappa)^{-1/2} \sqrt{h^2+\gamma _x \gamma _y+\left(\gamma _x+\gamma _y\right) \varepsilon _0 \pm A}, \quad\\
A&=&\sqrt{\left(h^2+\gamma _x^2+2 \gamma _x \varepsilon _0\right) \left(h^2+\gamma_y^2+2 \gamma _y \varepsilon _0\right)}.
\end{eqnarray}
%
%

The four roots of $Q$, $\pm \mathit{r}_{\pm}$, are branch points of $G_0$. The integrated density of states in the thermodynamic limit, $\mathcal{N}_0 (\varepsilon_0)$, now reads
%
%
\begin{eqnarray}
\label{eq:N_0}
\mathcal{N}_0 (\varepsilon_0)&=&\lim_{s\rightarrow \infty} \mathcal{N} (\varepsilon) =\lim_{s\rightarrow \infty} \mathcal{I} (\varepsilon) = \mathcal{I}_0 (\varepsilon_0)\\
&=&
\frac{1}{2 \rmi \pi}
\int_{\mathcal{C}_1} {\rm d} \alpha \: \big[ G^+_0 (\alpha)-G^-_0 (\alpha)\big] .
\end{eqnarray}
%
%

 A natural choice for the $G_0$ branch cuts is given by the curves $\mathcal{C}_0$ and $\mathcal{C}_1$, on which the $G$ poles accumulate as $s$ increases. It indeed corresponds to the direction, in the complex plane, for which the quantity computed in Eq.~(\ref{eq:N_0}) is real at each (infinitesimal) step of the integration. This latter condition was in fact implemented to draw the curves $\mathcal{C}_0$ and $\mathcal{C}_1$ in the different figures.

In the next section, we analyze in detail the four above-mentioned different regions in the phase diagram, in terms of $\mathcal{N}_0 (\varepsilon_0)$, its derivative, and the density of states  $\rho_0(\epsilon_0)=\partial_{\varepsilon_0}\mathcal{N}_0(\varepsilon_0)$. These quantities are, in most cases, computed as indicated in Eq.~(\ref{eq:N_0}). It may happen, as noted below, that the $\mathcal{C}_1$ curve has a complex shape, while $\mathcal{C}_0$ is simple. Since the integral over all branch cuts, corresponding to $\mathcal{C}_0$ and to $\mathcal{C}_1$, sums to unity, we can safely consider the integral over $\mathcal{C}_0$, instead of the nontrivial one over $\mathcal{C}_1$, and write $\mathcal{N}_0(\varepsilon _0)$ as one minus this integral. 
We also face the case of state degeneracies, with corresponding symmetric or nonsymmetric classical orbits. Each such orbit is considered separately by imposing the analyticity of $G_0$ in the region containing this orbit, bounded eventually by a closed branch cut on the sphere. The related $\Psi(\alpha)$ is zero along this line and can be considered as vanishing outside the considered region. This corresponds quite well to the (numerically derived) eigenstate  in the nonsymmetric case. However, in the symmetric case this description fails to reproduce the exact eigenstates since the latter is generically a linear combination of states located close to the classical orbits. 

%
\subsubsection {Analytical expressions of the densities of states}
%

A precise study of the branch cuts $\mathcal{C}_0$ and $\mathcal{C}_1$ allows one to distinguish between 
five different forms of the density of states (labeled $(a)$, $(b)$, $(c)$, $(d)$, and $(e)$ below) that can be expressed in terms of a complete elliptic integral of the first kind,
%
%
\begin{equation}
K(m)= \int_0^{\pi/2} (1-m \sin^2 \theta)^{-1/2} \mathrm{d} \theta,
\end{equation}
%
%
an incomplete elliptic integral of the third kind,
%
%
\begin{equation}
\Pi(n,\phi | m )= \int_0^{\phi} (1-n \sin^2 \theta)^{-1} (1-m \sin^2\theta)^{-1/2} \mathrm{d} \theta ,
\end{equation}
%
%
and a complete elliptic integral of the third kind, $\Pi(n|m)=\Pi(n,\pi/2 | m )$. 

Depending on the Hamiltonian parameters, we have already distinguished between four different zones, following the classical surface singularities. We will now show how these zones are characterized in terms of the density-of-states behavior. Indeed, each time a classical surface singularity (maximum, minimum, or saddle point) is crossed, the level sets (classical orbits or Husimi function local maxima) experience topological changes, as well as the integration contours, leading to a new expression for the integrated density of states. We now detail these different expressions, by describing each zone.

$\bullet$ Zone I:   $|\gamma_y | < \gamma_x <  h $.

Within this range of parameters (which coincides to the ``symmetric phase'' discussed in Sec.~\ref{subsec:PD}) the spectrum lies in the interval $-h \leqslant \varepsilon_0 \leqslant h$ and the density of states is a smooth decreasing function of the energy as can be seen in Fig.~\ref{fig:phase_diagram}.
The distribution of Majorana polynomial roots in this zone is similar to that displayed in Fig.~\ref{fig:Majsphere}$(b)$. In the complex plane, $\mathcal{C}_0$ and $\mathcal{C}_1$ lie in the imaginary and real axes  respectively. The integrated density of states is given by
%
%
\begin{eqnarray}
\label{eq:Int_dens_stat_b}
\mathcal{N}_0^{(b)}(\varepsilon _0) &=&
1+
\frac{
\sqrt{2}} {\pi  \mathit{r}_+ \sqrt{ - \kappa \ \gamma _x  \gamma _y }}
\Big[ a_-^2 \Pi \Big( \mu  \mathit{r}_-^2 \Big| \frac{\mathit{r}_-^2}{\mathit{r}_+^2} \Big) -
 \\
&&
a_+^2 \Pi \Big(\frac{\mathit{r}_-^2}{\mu } \Big| \frac{\mathit{r}_-^2}{\mathit{r}_+^2} \Big)+2 \sqrt{\gamma _x \gamma _y} (h+\varepsilon_0 )  K\left(\frac{\mathit{r}_-^2}{\mathit{r}_+^2}\right) \Big] \nonumber,
\end{eqnarray}
%
%
with
%
%
\begin{equation}
\label{eq:Int_dens_aux}
a_\pm  =  h \pm \sqrt{\gamma _x \gamma _y} \quad , \quad
\mu  =  \frac{\sqrt{\gamma _x}-\sqrt{\gamma _y}}{\sqrt{\gamma _x}+\sqrt{\gamma _y}}.
\end{equation}
%
%

$\bullet$ Zone II:  $ \left|\gamma_y \right| < h < \gamma_x $.

In this region, one must distinguish between two cases

$-$ II $(a)$: $- \frac{h^2 + \gamma_x^2}{2 \gamma_x}  \leqslant \varepsilon_0 \leqslant -h$.
$\mathcal{C}_0$ coincides with the whole imaginary axis while  $\mathcal{C}_1$ is made of two disconnected segments in the real axis as depicted in Fig.~\ref{fig:Majsphere}$(a)$. 
Here, the integrated density of states reads
%
%
\begin{eqnarray} 
\label{eq:Int_dens_stat_a}  
\mathcal{N}_0^{(a)}(\varepsilon _0)  &=& 1 + 
\frac{\sqrt{\kappa } \mathit{r}_+^2}  {\pi  \mathit{r}_- \sqrt{2 \gamma _x \gamma _y}   }
\bigg[ \\
&& \Pi \left(1-\frac{\mathit{r}_+^2}{\mu } \Big| 1-\frac{\mathit{r}_+^2}{\mathit{r}_-^2}\right)
   \left(1 -\frac{\mathit{r}_-^2}{\mu }\right) - \nonumber \\
&& \Pi \left(1-\mu  \mathit{r}_+^2 \Big| 1-\frac{\mathit{r}_+^2}{\mathit{r}_-^2}\right)
(1- \mu  \mathit{r}_-^2)
\bigg]. \nonumber 
\end{eqnarray}
%
%

$-$ II $(b)$: $-h \leqslant \varepsilon_0 \leqslant h$.
$\mathcal{C}_0$ and  $\mathcal{C}_1$ are the same as in zone I and the analytic expression of the density of states is given by Eq.~(\ref{eq:Int_dens_stat_b}).

These two branches $(a)$ and $(b)$ of the density of states diverge at  $\varepsilon_0=-h$. Indeed, the integrated density of states can be simplified into the form
%
%
\begin{eqnarray}
\label{eq:NII}
\mathcal{N}^{\rm II}_0(-h)  &=& 1+ \frac{2}{\pi \sqrt{\gamma _x \gamma _y}}  \bigg\{ \\
&& a_- \tan ^{-1} \bigg[\frac{a_-}{b_+(h)} \bigg]  - a_+ \tan ^{-1} \bigg[\frac{a_+}{b_0(h)}\bigg]   \bigg\} ,\nonumber
 \end{eqnarray}
%
%
with
%
%
\begin{eqnarray}
\label{eq:auxNII}
\hspace{-5mm} b_\pm(h) &=& \pm \big( \sqrt{h \gamma _x}-\sqrt{h \gamma _y} \big) +\sqrt{\left(\gamma _x-h\right) \left(h-\gamma _y\right)} , \\
\hspace{-5mm} b_0(h)&=&\sqrt{h \gamma _x} + \sqrt{h \gamma _y}+\sqrt{(\gamma _x-h)(h-\gamma_y)},
 \end{eqnarray}
%
%
and one can check that $\rho_0^{\rm II}(-h)=\partial_{\varepsilon _0} \mathcal{N}^{\rm II}_0(\varepsilon _0)|_{-h}$ diverges. One can further extract the leading behavior of the density of states near this point to obtain
%
%
\begin{equation}
\label{eq:density_of_states_EP-}
\lim_{\varepsilon _0\to -h} \rho _0^{\rm II}(\varepsilon_0)= -\frac{\log \left|  \varepsilon_0 + h  \right|}{2 \pi  \sqrt{\left(\gamma _x-h\right) \left(h-\gamma _y\right)}}.
 \end{equation}
%
%

$\bullet$ Zone III:  $  h < - \gamma_y  < \gamma_x $. 

In this region, one must distinguish between three cases

$-$ III $(a)$: $ - \frac{h^2 + \gamma_x^2}{2 \gamma_x} \leqslant \varepsilon_0 \leqslant -h $.
$\mathcal{C}_0$ and  $\mathcal{C}_1$ are the same as in II$(a)$, and the integrated density of states is given by Eq.~(\ref{eq:Int_dens_stat_a}).

$-$ III $(b)$: $-h \leqslant \varepsilon_0 \leqslant h$. 
$\mathcal{C}_0$ and  $\mathcal{C}_1$ are the same as in I, and the density of states $\mathcal{N}_0^{(b)}(\varepsilon _0)$ is given in Eq.~(\ref{eq:Int_dens_stat_b}).

$-$ III $(c)$: $ h \leqslant \varepsilon_0 \leqslant - \frac{h^2 + \gamma_y^2}{2 \gamma_y} $. 
$\mathcal{C}_0$ is made of two disconnected segments on the imaginary axis while $\mathcal{C}_1$ coincides with the whole real axis as depicted on the Majorana sphere  in Fig.~\ref{fig:Majsphere}$(c)$. The integrated density of states simply reads
%
%
\begin{eqnarray}
\label{eq:Int_dens_stat_c}
\mathcal{N}_0^{(c)}(\varepsilon _0)  & = & 1 - \mathcal{N}_0^{(a)}(\varepsilon _0),
\end{eqnarray}
%
%
where $\mathcal{N}_0^{(a)}$ is given in Eq.~(\ref{eq:Int_dens_stat_a}). 

In this zone III, the density of states has two singularities at $\varepsilon_0=\pm h$. The integrated density of states for these energies is given by $\mathcal{N}^{\rm III}_0(-h)=\mathcal{N}^{\rm II}_0(-h)$ [see Eq.~(\ref{eq:NII})] 
and
%
%
\begin{equation}
\label{eq:NIII}
\mathcal{N}^{\rm III}_0(h)  = \frac{2}{\pi 
   \sqrt{\gamma _x \gamma _y}} \bigg[ a_+ \tan ^{-1} \frac{a_+}{b_0(-h)}  - a_- \tan ^{-1} \frac{a_-}{b_-(-h)}   \bigg].
 \end{equation}
%
%

As done in zone II, one can compute  the leading behavior of the density of states near these points and one gets
%
%
\begin{equation}
\label{eq:density_of_states_EP+}
\lim_{\varepsilon _0\to +h} \rho _0^{\rm III}(\varepsilon_0)= -\frac{\log \left|  \varepsilon_0 - h  \right|}{2 \pi  \sqrt{-\left(\gamma _x+h\right) \left(h+\gamma _y\right)}}.
 \end{equation}
%
%

For $\gamma_x=-\gamma_y$, the spectrum is symmetric with respect to $\varepsilon_0=0$ and the above expression gives the exact location, in the thermodynamic limit,  of the so-called exceptional point observed in 
Refs.~\cite{Heiss05,Heiss06} where a more complex diverging behavior was conjectured.

$\bullet$ Zone IV: $ h  <  \gamma_y  < \gamma_x $. 

In this zone the density of states presents three different regions, of type $(d)$, $(e)$, and $(b)$.  The curve $\mathcal{C}_1$ is more complex here, while $\mathcal{C}_0$ always lies on a straight line in the complex plane. This is why we choose to integrate around $\mathcal{C}_0$ instead of $\mathcal{C}_1$. 

\begin{figure}[t]
 \centering
\includegraphics[width=8.5cm]{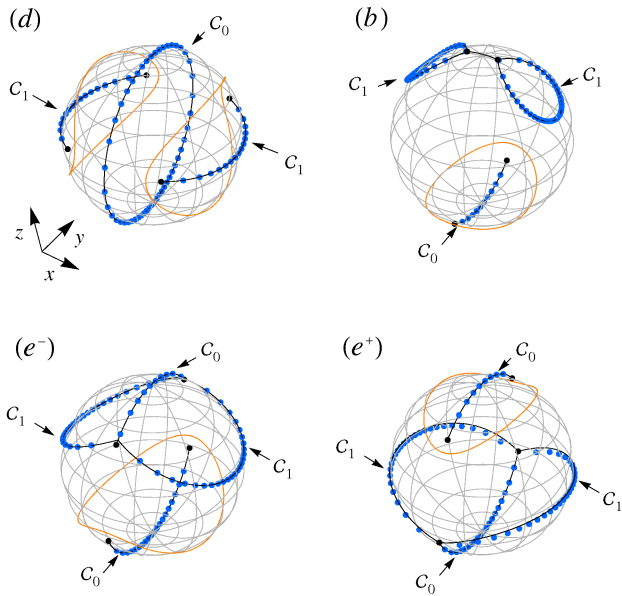}
\caption{\label{fig:OrbitZIV} Roots of the Majorana polynomial (blue dots) $(\gamma_x =10, \gamma_y=5, h= 1$, and $s=40$), classical orbits (orange curves), $\mathcal{C}_0$ and $\mathcal{C}_1$ (black curves), for eigenstates (labelled by $n$) in zone IV$(d)$ ($n=15$), zone IV$(e)$ [$(e^-)$: $n=25$, $(e^+)$: $n=26$ ] and zone IV$(b)$ ($n=35$).
In zone IV$(e)$ two kinds of states coexist, of type  $(e^-)$ and $(e^+)$, associated with the two classical orbits nonrelated by symmetry that alternate in the spectrum.}
\end{figure}

$-$ IV$(d)$: $- \frac{h^2 + \gamma_x^2}{2 \gamma_x} \leqslant \varepsilon_0 \leqslant - \frac{h^2 + \gamma_y^2}{2 \gamma_y}$. $\mathcal{C}_0$ coincides with the whole imaginary axis while $\mathcal{C}_1$ has two disconnected branches lying symmetrically on the unit circle with respect to the imaginary axes. 
We are here facing a case where the classical orbits are related by symmetry [see Fig.~\ref{fig:OrbitZIV}$(d)$]. 
One finds, for this region,
%
%
\begin{eqnarray}
\label{eq:Int_dens_stat_d}
\mathcal{N}_0^{(d)}(\varepsilon _0)  &=& 1+
\frac{2 \sqrt{2} \mathit{r}_-} {\pi \left(\mathit{r}_- -\mathit{r}_+\right) \sqrt{-\gamma _x \gamma _y \left(h+\varepsilon _0\right)}}
\bigg[ \nonumber \\
&& \frac{ a_-^2}{u\left(-\mathit{r}_-\right) u\left(\mathit{r}_-\right)} \mathcal{E}\left(\mathit{r}_-,y\right) -\nonumber \\
&& \frac{a_+^2 }{\mathit{r}_-^2 u\left(-\frac{1}{\mathit{r}_-}\right) u\left(\frac{1}{\mathit{r}_-}\right)}\mathcal{E}\left(\frac{1}{\mathit{r}_-},-y\right) \bigg],
\end{eqnarray}
%
%
with 
%
%
\begin{eqnarray}
\label{eq:Elli}  
\mathcal{E}\left(\mathit{r}_-,y\right) &= & 
\Pi \left[-\frac{u\left(-\mathit{r}_-\right)}{y u\left(\mathit{r}_-\right)},\sin^{-1}\sqrt{-y} \Big| \frac{1}{y^2}\right] - \nonumber \\
&&\Pi \left[-\frac{u\left(-\mathit{r}_-\right)}{y u\left(\mathit{r}_-\right)},\sin ^{-1}\sqrt{y} \Big| \frac{1}{y^2}\right]- \nonumber \\
&& \Pi \left[-\frac{u\left(\mathit{r}_-\right)}{y u\left(-\mathit{r}_-\right)},\sin^{-1}\sqrt{-y} \Big| \frac{1}{y^2}\right]+ 
\nonumber \\&&\Pi \left[-\frac{u\left(\mathit{r}_-\right)}{y u\left(-\mathit{r}_-\right)},\sin ^{-1}\sqrt{y} \Big| \frac{1}{y^2}\right],
\end{eqnarray}
%
%
where
%
%
\begin{eqnarray}
\label{eq:Elli_aux} 
y &=& \frac{\mathit{r}_--\mathit{r}_+}{\mathit{r}_-+\mathit{r}_+} , \\
u\left(\mathit{r}_-\right) &=& \sqrt{\sqrt{\gamma _x}-\sqrt{\gamma _y}} \mathit{r}_-+\sqrt{\sqrt{\gamma _x}+\sqrt{\gamma _y}}.
\end{eqnarray}
%
%
 
$-$ IV$(e)$: $ - \frac{h^2 + \gamma_y^2}{2 \gamma_y} \leqslant \varepsilon_0 \leqslant -h$. This region shows two disconnected classical trajectories not related by symmetry (see Fig.~\ref{fig:OrbitZIV}), corresponding to two qualitatively different kinds of states which alternate in the spectrum. 
$\mathcal{C}_0$ comprises two disconnected components lying in the imaginary axis, while $\mathcal{C}_1$ is still complex and, moreover, is different for the two kinds of states. One finds
%
%
\begin{eqnarray}
\label{eq:Int_dens_stat_e} 
\mathcal{N}_0^{(e)}(\varepsilon _0)   &=&    1 + \frac{\sqrt{2} }{\pi  \mathit{r}_+
   \sqrt{-\kappa  \gamma _x \gamma _y}} 
\bigg\{
-4 \varepsilon_0 \sqrt{\gamma _x \gamma _y} K\left(\frac{\mathit{r}_-^2}{\mathit{r}_+^2}\right) + \nonumber\\
&&
 a_-^2 \bigg[  \Pi \left(\frac{1}{\mu  \mathit{r}_+^2} \Big| \frac{\mathit{r}_-^2}{\mathit{r}_+^2}\right)
-\Pi\left(\mu  \mathit{r}_-^2 \Big| \frac{\mathit{r}_-^2}{\mathit{r}_+^2}\right)  \bigg]  + \nonumber\\
&&
a_+^2 \bigg[ \Pi\left(\frac{\mathit{r}_-^2}{\mu } \Big| \frac{\mathit{r}_-^2}{\mathit{r}_+^2}\right)
- \Pi \left(\frac{\mu}{\mathit{r}_+^2} \Big| \frac{\mathit{r}_-^2}{\mathit{r}_+^2}\right) \bigg]  
\bigg\} .
\end{eqnarray}
%
%

For the critical energy, at the boundary between IV$(d)$ and IV$(e)$, the integrated density of states simplifies to
%
%
\begin{equation}
\label{eq:NIVa} 
\mathcal{N}^{\rm IV}_0 \bigg( - \frac{h^2 + \gamma_y^2}{2 \gamma_y} \bigg) = 1 + \frac{1}{\pi  \sqrt{\gamma _x \gamma _y}}  \big[  a_-  c(-h) -a_+ c(h)\big] ,
\end{equation}
%
%
with
%
%
\begin{eqnarray}
\label{eq:NIVb}
c(h)&=& \tan^{-1}\left [\frac{h \sqrt{\gamma_x}+\gamma _y^{3/2}}
{\sqrt{(\gamma _x -\gamma _y)(\gamma_y^2-h^2)}}\right] , \\
\mathcal{N}^{\rm IV}_0 (-h)  &=&1 - \frac{h}{\sqrt{\gamma _x\gamma _y}}.
 \end{eqnarray}
%
%

In addition, the density-of-states singular behavior is not symmetrical and reads
%
%
\begin{eqnarray}
\label{eq:density_of_states_EP--}
\hspace{-5mm} \lim_{\varepsilon _0\to \left( -\frac{ h^2 + \gamma_y^2}{2 \gamma _y} \right)^-}\rho _0^{(e)}
&=& -\frac{\log \left|  \varepsilon_0 + \frac{ h^2 + \gamma_y^2}{2 \gamma _y}  \right| \sqrt{\gamma _y}}{\pi  \sqrt{\left(\gamma _x-\gamma _y\right) \left(\gamma _y^2-h^2\right)}} ,\\
&=& 2 \lim_{\varepsilon _0\to \left( -\frac{ h^2 + \gamma_y^2}{2 \gamma _y} \right)^+} \rho _0^{(d)} .
 \end{eqnarray}
%
%

$-$ IV$(b)$: $-h \leqslant \varepsilon_0 \leqslant h$.  $\mathcal{C}_0$ is simply connected and lies on the imaginary axes. Like in the previous case, $\mathcal{C}_1$ is nontrivial 
(see Fig.~\ref{fig:OrbitZIV}). Nevertheless, the expression found for  $\mathcal{N}_0$ in this region coincides with that given by Eq.~(\ref{eq:Int_dens_stat_b}). 

%
%
%
%

 We now discuss the particular features found in the spectral region IV$(e)$. At $\varepsilon_0 = -h$,  the density of states is discontinuous (see Fig.~\ref{fig:phase_diagram}), a fact which can be understood already from the topological analysis of the classical surface $\mathrm{H}_{0}$. Indeed, the transition from zone $(e)$ to zone $(b) $ corresponds to leaving a local maximum of $\mathrm{H}_{0}$ (see 
Fig.~\ref{fig:mean_field_surface}); therefore, a family of classical orbits no longer contributes to the density of states.

\begin{figure}[t]
  \centering
\includegraphics[width=7cm]{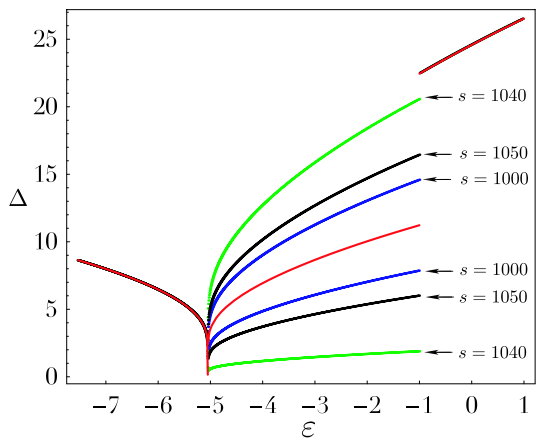}
  \caption{Gap between two consecutive levels as a function of the energy in region IV for 
  $\gamma_x=15$, $\gamma_y=10$ and $h=1$. In the central region, one sees a real lack of convergence toward the red line when increasing $s$, which is the average gap as computed in the thermodynamic limit.}
  \label{fig:gap}
\end{figure}

In addition, as opposed to all other regions, the energy difference between two consecutive levels, $\Delta^{(i)}=E^{(i+1)}-E^{(i)}$, computed for increasing $s$, does not converge towards the analytical result and, actually, does not converge at all. In region IV$(e)$, $\Delta^{(i)}$ spreads over two branches $(+)$ and $(-)$, depending on the parity of the $i$, which oscillate without converging as $s$ increases, as can be seen in 
Fig.~\ref{fig:gap}. In this case, the gap we compute, in the thermodynamic limit, is actually the average gap, namely $\Delta_0(\varepsilon_0)=\frac{1}{2}\big[ \Delta^{(+)}(\varepsilon_0)+\Delta^{(-)}(\varepsilon_0)\big]$.
This is clearly to be understood in relation to the existence of two kinds of states alternating in the spectrum. Indeed, when analyzed separately within each set of states ($e^+$ or $e^-$), the computed energy gaps (between levels $j$ and $j+2$ in the energy spectrum) converge as $s\to\infty$. In addition, both such gaps converge to twice the value of $\Delta_0(\varepsilon_0)$ (otherwise the two kind of states would not alternate as observed numerically). The oscillatory behavior  noted in Fig.~\ref{fig:gap} signals an energy drift (with $s$) of one set of energy levels with respect to the other.
%
%
%

%
%
\section{Finite-size corrections} 
\label{sec:finite}
%
%

In the previous section, we have analyzed the thermodynamic limit of the LMG model spectrum by considering the leading terms in the expansion (\ref{eq:one_over_s_development}) [order $(1/s)^0$]. We now express the next-order corrections, which have already been shown, at least for the ground state, to display nontrivial scaling properties \cite{Botet83,Dusuel04_3,Dusuel05_2,Leyvraz05}. For the sake of simplicity, we limit the present analysis to the case $\gamma_x=1$, and $\gamma_y=0$. 

%
%
\subsection{   First-order expansion for \textit{G}  }
%
%

Identifying terms of order $1/s$ in Eq.~(\ref{eq:G_Schrodinger}), one obtains the following form for the first-order term of $G$:
%
%
\begin{eqnarray}
\label{eq:G1}
G_1^\pm (\alpha) = \hat{G}_1 (\alpha) + \tilde{G}_1^\pm (\alpha) ,
\end{eqnarray}
%
%
with
%
%
\begin{eqnarray}
\label{eq:G1hat_et_tilde}
\hat{G}_1 (\alpha) &=& \frac{h \alpha  \left[h \left(\alpha ^2+1\right)  -\alpha ^2+1 \right]}{2 \left(1-\alpha ^2\right)  Q(\alpha)}, \\
\tilde{G}_1^\pm (\alpha) &=&  \pm\frac{h \left(\alpha ^2+1\right)+2 \left(\alpha ^2-1\right) \varepsilon _1}{2 \left(\alpha ^2-1\right) \sqrt{2 Q(\alpha)}}.
\end{eqnarray}
%
%
$\hat{G}_1$ is thus an analytic function of $\alpha$ with poles at $\pm \mathit{r}_-$ and $\pm \mathit{r}_+$ while 
$\tilde{G}_1$ has the same branch cuts as $G_0$. 
$\mathcal{I}(\varepsilon)$ reads, recalling Eq.~(\ref{eq:int_of_G}) and developing up to first order,
%
%
\begin{eqnarray}
\mathcal{I}(\varepsilon)  &=& \frac{1}{2 \rmi \pi} 
\oint_{\widetilde{\mathcal{C}}_1} G_0(\alpha)  \:  {\rm d} \alpha   + 
\frac{1}{s}  \frac{1}{2 \rmi \pi}  \oint_{\widetilde{\mathcal{C}}_1} G_1(\alpha)  \:  {\rm d} \alpha , \\
&=& \mathcal{I}_0(\varepsilon) + \frac{1}{s} \mathcal{I}_1(\varepsilon),
\end{eqnarray}
%
%
where $\mathcal{I}_0(\varepsilon)$ is given in Eq.~(\ref{eq:N_0}) and where one can rewrite
%
%
\begin{equation}
\mathcal{I}_1(\varepsilon) =  \frac{1}{4}+\frac{1}{2 \rmi \pi}
\int_{\mathcal{C}_1}  {\rm d} \alpha \: \big[ \tilde{G}^+_1 (\alpha)-\tilde{G}^-_1 (\alpha)\big] ,
\end{equation}
%
%
 the $\frac{1}{4}$ coming from the integration over the poles.

For $ \gamma_x=1,  \gamma_y=0$, one has only zones I and II to consider, which focuses the analysis on only two energy regions. In zones I and II$(b)$ one obtains
%
%
\begin{equation}
\mathcal{I}_1^{(b)}(\varepsilon)  =   \frac{1}{4}  + \frac{  (h +2 \varepsilon
   _1) K\left(\frac{\mathit{r}_-^2}{\mathit{r}_+^2}\right)-2 h \ \Pi
   \left(\mathit{r}_-^2 \Big| \frac{\mathit{r}_-^2}{\mathit{r}_+^2}\right) }{\pi  \mathit{r}_+ \sqrt{-\kappa  }} ,
\end{equation}   
%
%
whereas in region II$(a)$ one finds
%
%
\begin{eqnarray}
\mathcal{I}_1^{(a)}(\varepsilon)  &=&  \frac{1}{\pi  \sqrt{\kappa }} \Bigg\{ 
\frac{2 h}{\mathit{r}_- \left(\mathit{r}_+^2-1\right)} \nonumber\\
&& 
\Bigg[ K\left(1-\frac{\mathit{r}_+^2}{\mathit{r}_-^2}\right) -\mathit{r}_+^2 \Pi \left(1-\mathit{r}_+^2 \Big| 1-\frac{\mathit{r}_+^2}{\mathit{r}_-^2}\right)  \Bigg] +\nonumber\\
&& \frac{h+2 \varepsilon _1}{\mathit{r}_+} K\left(1-\frac{\mathit{r}_-^2}{\mathit{r}_+^2}\right)  \Bigg\} .
\end{eqnarray}
%
%

\begin{figure}[t]
 \centering
\includegraphics[width=\columnwidth]{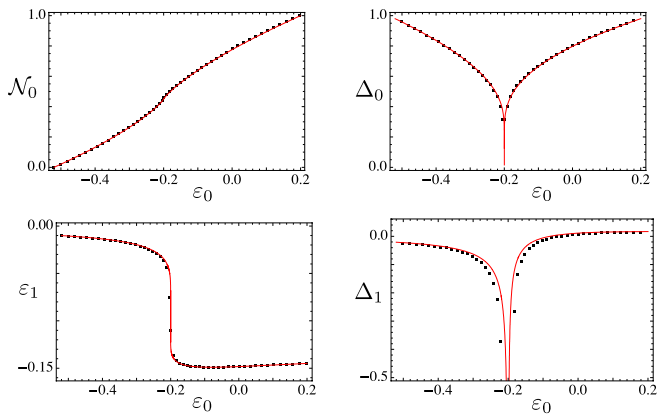}
\caption{\label{fig:FigOneOverS} Comparison between analytical (red line) and numerical ($s=50$ black dots) results for the (zeroth-order) integrated density of states $\mathcal{N}_0$ (upper left) and energy gap $\Delta_0$ (upper right) and the first-order finite-size corrections to the energy $\varepsilon_1$ and to the gap ($\Delta_1$, lower right).}
\end{figure}

Now, for all $s$, we expect that $\mathcal{I}(\varepsilon)=\mathcal{I}_0(\varepsilon_0)$, which implies, at order $1/s$, $\mathcal{I}_1(\varepsilon-\varepsilon_1/s)=\mathcal{I}_1(\varepsilon_0)=0$. This condition allows one to compute the first-order correction to the energy, $\varepsilon_1$, which is displayed in 
Fig.~\ref{fig:FigOneOverS} (lower left) and compares nicely with the numerical values, already for small values of $s$ (here $s=50$).

%
%
%
%

\subsection{Energy gaps}

The gap between two successive levels has already been  discussed above in the zone-IV case. At the thermodynamic limit, it generically reads

%
%
\begin{equation}
\Delta_0(\varepsilon_0)=\frac{1}{\rho_0(\varepsilon)}=\frac{\partial \varepsilon_0}{\partial \mathcal{N}_0(\varepsilon_0)}. 
\end{equation}
%
%
 
With the analysis done in the previous section, we can now compute finite size corrections to the gap. To first order, we obtain 

%
\begin{equation}
\Delta  =  \Delta_0  + \frac{1}{s} \Delta_1
           =   \Delta_0 \bigg( 1 + \frac{1}{s} \frac{\partial \varepsilon _1}{\partial \varepsilon_0} \bigg) .
\end{equation}
%
%

The above derived values of $\varepsilon _1$ allow us to get a closed form for $\Delta_1$, which nicely compares to the numerical values, as can be seen in Fig.~\ref{fig:FigOneOverS} (lower right) for $s=50$.

The  $\Delta_1$ correction is singular at the exceptional points, which are, as discussed in Sec.~\ref{sec:thermo}, located at $\varepsilon_0=-h$. Note that Leyvraz and Heiss numerically found a logarithmic singularity at the exceptional points \cite{Leyvraz05}. A related feature was already observed for the gap between the ground state and the first-excited state \cite{Dusuel04_3,Dusuel05_2}. In the latter case, a  scaling hypothesis led to a derivation of the first-order correction, showing a $N^{-1/3}$ behavior.
Unfortunately, the scaling hypothesis cannot be used here at the exceptional points. We have determined the behavior of the gap in their vicinity; setting $\eta = | h +\varepsilon_0 |$, one gets
%
%
\begin{eqnarray}
\Delta (\varepsilon_0 \to - h^+  )& = & -\frac{2\pi \sqrt{(1-h) h}}{\log \eta } 
\bigg\{ 1 -     \frac{1}{s} \bigg[ \frac{1}{4(h-1)}+  \nonumber \\
&& \frac{\sqrt{(1-h) h} \sin^{-1}(1-2 h)}{\eta  \log^2 \eta} \bigg]     \bigg\} , \\
\Delta (\varepsilon_0 \to - h^-  )& = & -\frac{2 \pi \sqrt{(1-h) h} }{\log \eta} \bigg[ 1 -     \frac{1}{s}   \nonumber \\
&& \frac{2 \sqrt{(1-h) h} \sin ^{-1}\sqrt{h}}{\eta  \log ^2 \eta}   \bigg].
\end{eqnarray}
%
%
Note that the leading term is simply the inverse of $\rho_0$, which is given in Eq.~(\ref{eq:density_of_states_EP-}) and vanishes when $\eta$ goes to zero.

%
\section{Observable expectation values}
\label{sec:observables}
%
In this section, we discuss the expectation values of spin observables for generic eigenstates of the LMG model. 
The simplest way to perform such a calculation is to use the Hellmann-Feynman theorem, which relates these expectation values to the partial derivative of the eigenenergies with respect to Hamiltonian parameters. For instance

\begin{equation}
\label{eq:Hell-Feyn-0}
\bra{\Psi} S_z \ket{\Psi} = - \partial_h E \quad , \quad
\bra{\Psi} S_{x}^2 \ket{\Psi} = - 2 s \ \partial_{\gamma_{x}} E.
 \end{equation}
 
\begin{figure}[t]
\center
\includegraphics[width= \columnwidth]{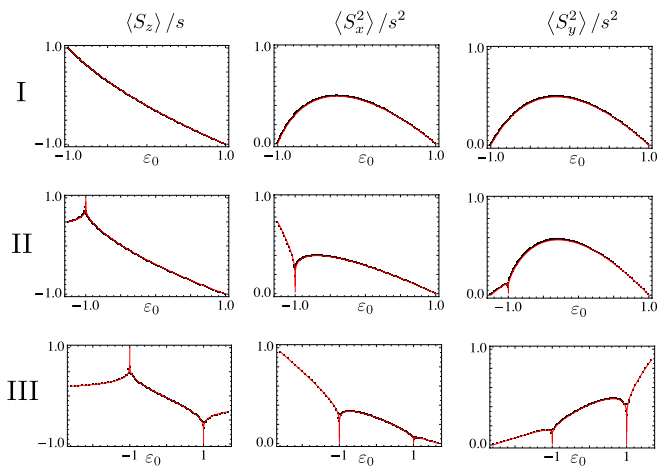}
\caption{\label{fig:ObsZI} Comparison of expectation values of several observables obtained from numerical diagonalizations (black dots) and from the Hellmann-Feynman theorem in the thermodynamic limit (red lines). Plot parameters: $s=60$, zone $ {\rm{I}}: (\gamma_x=1/2, \gamma_y=1/3,h= 1)$, zone $ {\rm{II}}: (\gamma_x=2, \gamma_y=1/2,h= 1)$, and zone $ {\rm{III}}: (\gamma_x=5, \gamma_y=-3,h=1 )$.}
\end{figure}

\begin{figure}[t]
 \centering
\includegraphics[width= \columnwidth]{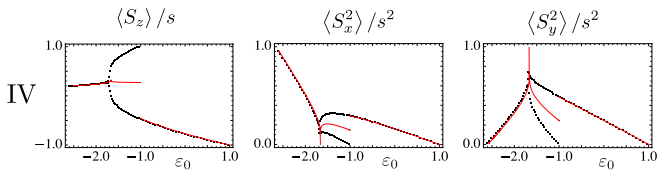}
\caption{\label{fig:ObsZIV} Same as Fig.~\ref{fig:ObsZI}, for a typical point in zone IV $(\gamma_x=5, \gamma_y=3,h=1 )$ and $s=60$. 
In the central region [zone IV$(e)$], there is a clear discrepancy between the numerical values (black dots) and those derived from the Hellmann-Feynman theorem (red lines).}
\end{figure}

As an illustration, we compare in Figs.~\ref{fig:ObsZI} and \ref{fig:ObsZIV} three cases, computed numerically (at finite $s$) and via the Hellmann-Feynman theorem in the thermodynamic limit, {\it i.~e.} replacing $E$ by 
$s \, \varepsilon_0$. As expected, one can see an almost perfect agreement, except for zone IV$(e)$ discussed below.
%
%

%
%

Let us still make use of the semiclassical analysis discussed in previous sections. The expectation value $ \bra{\Psi} \hat{O} \ket{\Psi}$ for an observable $O$  reads \cite{Paul93}, at leading order,
%
%
\begin{equation}
\label{eq:matrix_element_ell_3}
\langle \hat{O} \rangle = \frac{ \bra{\Psi} \hat{O} \ket{\Psi}}{\braket{\Psi}{\Psi}} =  
\frac{1}{T} \int_0^{T} \mathrm{d} t  \:  \bra{\alpha(t)}\hat{O} \ket{\alpha(t) } ,
\end{equation}
%
%
where $T$ is the period of the classical orbit with energy $\varepsilon_0$ and $\alpha(t)$ the solution of the classical dynamics equation \cite{Kurchan89}. 

Let us focus on the $ \langle S_z \rangle$ case.
In zone I, it is maximal for the ground state. Indeed, in that region, ${\mathrm H}_0$ is minimum for $\alpha=0$, where the classical orbit degenerates to a single point at which the ground-state amplitude $|\Psi(\alpha)|^2$ is concentrated. As a result, although this true ground state differs from the simple fully polarized state \cite{Orus08_2},  $ \langle S_z \rangle$ reaches its maximum value $s$. 

This also occurs in regions II and III, for energies corresponding to the exceptional points. Here, classical orbits display a  characteristic ``figure-8'' shape, with  the values of $\alpha$  therefore differing from zero. The saturation effect results in that case from the fact that the period of the orbit  diverges, with a  vanishingly small classical velocity near $\alpha=0$, forcing the expression in Eq.~(\ref{eq:matrix_element_ell_3}) to saturate. In all cases except zone IV$(e)$, this latter computation leads to the same result as that simply obtained from the Hellmann-Feynman theorem.

In zone IV$(e)$, the numerically computed expectation values alternate along two distinct curves, differing from the Hellmann-Feynman result. This corresponds to the already discussed existence, for the same cenergy $\varepsilon_0$, of two kinds of classical trajectories nonrelated by symmetry (see Fig.~\ref{fig:OrbitZIV}).
For each numerically derived eigenstate, the associated $|\Psi(\alpha)|^2$ concentrates alternatively near one of the two classical orbits. Integrating separately along each orbit precisely gives the two branches that are observed numerically (Fig.~\ref{fig:ObsZIV}), while the Hellmann-Feynman computation leads to an averaged value.

%
%
\section{Conclusion}
\label{sec:conclusion}

We have studied in detail the full spectrum of the Lipkin-Meshkov-Glick model by means of a coherent-states formalism. In a first step, we simply determined the main characteristics of the (zero temperature) phase diagram by analyzing extrema and saddle points of the classical energy surface. This leads us to distinguish between four zones in the phase diagram corresponding to various patterns of the density of states whereas the usual ground-state criterion leads to only two distinct phases.  

In a second step, we analyzed more deeply the nature of the eigenstates in terms of their associated Majorana polynomial roots. This enabled us to exactly compute the integrated density of states in the thermodynamic limit as well as the first finite-size corrections. This remarkable result mainly stems from the fact that the roots of the Majorana polynomial lies on well-defined curves, where their density varies monotoneously with the energy. We also clarified the nature of the so-called ``exceptional'' points in the spectrum. 

Finally, we addressed the question of computing generic observable expectation values, in particular when, owing to subtle spectral reasons, the Hellmann-Feynman theorem cannot be used. 

In principle, the same type of analysis could be performed for any spin Hamiltonian expressed in terms of single-spin operators (so-called ``collective models''). Preliminary investigations of such models with cubic or quartic interactions are currently under study. 
Another perspective, also presently under investigation, concerns the dynamical properties for evolutions under both fixed and variable Hamiltonian parameters. 

\acknowledgments

We are grateful to T. Paul for fruitful and stimulating discussions and to S. Garmon for a careful reading of the manuscript. P. R. was partially supported by FCT and EU FEDER through POCTI and the QuantLog POCI/MAT/55796/2004 Project of CLC-DM-IST, SQIG-IT and grant No. SFRH/BD/16182/2004/2ZB5.

\appendix

%
%
\section{Mapping the LMG model onto an equivalent one-dimensional model}
\label{sec:mapping}
%
%

The  density-of-states calculation given in this paper relies on the fact that the roots of the Majorana polynomial lie on well-defined curves in the complex plane. This result stems from the well-known wave-function node oscillation theorem for one-dimensional systems, which arise here via a mapping of the LMG model onto the problem of a particle in a one-dimensional potential (see  \cite{Ulyanov92} for a review), which we summarize here. A one-to-one relation exists between the energy spectrum of the spin system and the low-lying quantum states of such a particle. 

We aim to rewrite the equation for the eigenstate $\Psi(\alpha)$ as a Schr\"{o}dinger  equation for a particle moving in a one-dimensional potential. 
The procedure consists in three steps, given first for the case $ \gamma_y < 0$.

\begin{enumerate}

\item We change $H$ into an equivalent form such that the roots of the Majorana polynomials (nodes of the wave-function) which are aligned on the $\mathcal{C}_1$ curve are sent onto the unit circle. This is achieved through the following unitary transformation: $\tilde{H}= \mathrm{e}^{\mathrm{i} \frac{\pi}{2} S_x } H \mathrm{e}^{- \mathrm{i} \frac{\pi}{2} S_x }$

\item The unit circle being parametrized by an angle $\theta$, we write  $\Phi(\theta)=  \mathrm{e}^{- \mathrm{i} s \theta } \Psi( \mathrm{e}^{\mathrm{i} \theta })$ for $\theta \in [0,2 \pi [ $.

\item Finally, we define a new function $\phi(x)$, which satisfies a one-dimensional Sch\"{o}dinger equation and such that part of its spectrum is put in one-to-one correspondance with the original spin spectrum. This is achieved by setting $\Phi(\theta) = \mathrm{e}^{f[ x (\theta) ] } \phi[ x(\theta) ]  $ where $f(x) $ and $x(\theta)$ are chosen to suppress the first-order derivative in the initial Equation (\ref{eq:Maj_Schrodinger}) for $\Psi(\alpha)$ and to set the ``mass'' term equal to $s$. 
The resulting Schr\"{o}dinger-like equation for $ \phi(x) $, describing a particle in a one-dimensional periodic potential, reads
%
%
\begin{equation}
\label{eq:1D_Schr}
- \frac{1}{2 s} \partial_x^2\phi(x) + V(x) \phi(x)= E \phi(x).
\end{equation}
%
%

\end{enumerate}

Following this procedure, one obtains the effective potential
%
%
\begin{eqnarray}
\label{eq:1D_SchrV2}
V(x) &=& \frac{1}{2 \gamma _y-2\gamma _x \ \mathrm{sn} (B| \gamma _x / \gamma _y)^2} \Big\{ \nonumber \\
&&  h (2 s+1) \left(\gamma _x-\gamma _y\right)  \mathrm{sn}(B| \gamma _x / \gamma _y)-  \nonumber \\
&&  \left[h^2 s+(s+1) \gamma _x \gamma _y\right] \mathrm{cn}(B| \gamma _x / \gamma _y)^2 \Big\}
,
\end{eqnarray}
%
%
with
%
%
\begin{equation}
B= \sqrt{-\gamma _y} \ x+K\left(\frac{\gamma _x}{\gamma _y}\right).
\end{equation}
%
%
Note that $V$ is periodic with period $L = \frac{4}{\sqrt{-\gamma_y}} K\left(\frac{\gamma _x}{\gamma _y}\right)$. 

The mapping onto a one-dimensional potential and the celebrated node oscillation theorem allows one to sort the eigenstates of increasing energy according to their number of nodes. Clearly, a $\phi(x)$ node leads to a $\Psi(\alpha)$ node for the corresponding LMG eigenstate. The first $(2s+1)$ eigenstates of this Hamiltonian $\tilde{H}$ correspond to the eigenstates of the LMG Hamiltonian with the same energy. Note that, since we focus in this paper on the ($s+1$)-dimensional ``even-$m$'' sector, this leads eventually to a node number inceasing by steps of 2 for each new eigenstate.

%
%
\begin{figure}[th]
 \centering
\includegraphics[width= \columnwidth]{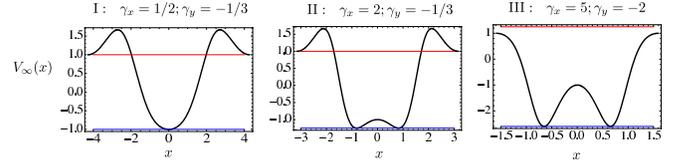}
\caption{\label{fig:1DPotentialm0} Effective one-dimensional potential in the thermodynamic limit $V_\infty(x)=\lim_{s\to\infty} \frac{V(x)}{s}$ for $\gamma_y<0$ and $h=1$. Blue and red lines are respectively the lower and upper bounds of the spin system spectrum $\varepsilon_0=\frac{E}{s}$.}
\end{figure}
%
%

Typical potentials are shown in Fig.~\ref{fig:1DPotentialm0}, with parameters associated with regions I, II and III of the LMG phase diagram. The LMG spectrum corresponds to the energies lying between the lower (blue) and the upper (red) lines. The qualitative differences between the three regions appear clearly here. Indeed, in region I the particle  moves in a single-well potential whereas it is in a double-well potential in region II.
In region III, a higher ``allowed'' energy region appears, with the extended (unbounded) states above the potential barrier. Crossing the latter corresponds to the upper density-of-states singularity discussed in the text. Note, however, that the extended or bounded nature of the eigenstates for this equivalent one-dimensional system does not have a direct translation into the nature of the corresponding eigenstates in the LMG problem.

\begin{figure}[t]
 \centering
\includegraphics[width=\columnwidth]{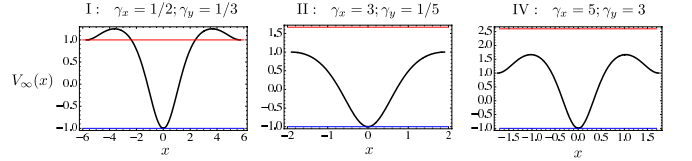}
\caption{\label{fig:1DPotentialp0} Effective one-dimensional potential in the thermodynamic limit $V_\infty(x)=\lim_{s\to\infty} \frac{V(x)}{s}$ for $\gamma_y>0$ and $h=1$. Blue and red lines correspond, respectively, to the upper and lower bounds of the energy 
$\varepsilon_0=\frac{E}{s} $ in the LMG problem.}
\end{figure}

Similar transformations can be achieved for positive $\gamma_y$ but in this case, one must consider 
$\tilde{H}= - \mathrm{e}^{\mathrm{i} \frac{\pi}{2} S_y } H \mathrm{e}^{- \mathrm{i} \frac{\pi}{2} S_y }$. Note the occurence of the minus sign which maps the high-energy states of the LMG model onto the low-energy states of the particle-problem (and reciprocally). Following steps (2) and (3), one obtains the potential

%
%
\begin{eqnarray}
\label{eq:1D_SchrV1} 
V(x) &=&\frac{1}{2 \gamma _y \mathrm{cn} [C \ |\gamma _y/(\gamma _y-\gamma _x)]^2-2 \gamma _x}
\bigg\{ \\
&& h (2 s+1) \left(\gamma _x-\gamma _y\right) \mathrm{cn}[C \ |\gamma _y/(\gamma _y-\gamma _x)]-\nonumber \\
&&\left(h^2 s+(s+1) \gamma _x \gamma _y\right) \mathrm{sn}[C \ |\gamma _y/(\gamma _y-\gamma _x)]^2 \bigg\}, \nonumber
\end{eqnarray}
%
%
with
%
%
\begin{equation}
C= \sqrt{\gamma _x-\gamma _y} \ x.
\end{equation}
%
%
Here, $V$ is periodic with period  $L =\frac{4}{\sqrt{\gamma _x-\gamma _y}} K\left(\frac{\gamma _y}{\gamma _y-\gamma _x}\right)$. 
The effective potentials are displayed on Fig.~\ref{fig:1DPotentialp0} for zones I, II and IV, where some care must now be taken for the correspondence with the LMG model. The upper levels (close to the upper red line) correspond to the lower levels in the LMG case. 

%
%
%
%


\end{document}